\documentclass{iau} 
\usepackage[T1]{fontenc}
\usepackage{graphicx}
\usepackage{natbib}
\usepackage{amssymb}
\usepackage{mathabx}

\title[White-dwarf asteroseismology] 
{White-dwarf asteroseismology: an update}

\author[]   
{Alejandro H.\ C\'orsico$^{1,2}$}

\affiliation{$^1$Facultad de Ciencias Astron\'omicas y Geof\'isicas, Universidad Nacional de La Plata, Paseodel Bosque s/n, (1900) La Plata, Argentina, \\
$^2$  Instituto de Astrof\'isica de La Plata, IALP (CCT La Plata), CONICET-UNLP\\
  email: {\tt acorsico@fcaglp.unlp.edu.ar} }

\pubyear{2019}
\volume{357}  
\jname{White Dwarfs as Probes of Fundamental Physics and Tracers of Planetary, Stellar, and Galactic Evolution}
\begin{document}

\maketitle

\begin{abstract}
The vast majority of stars that populate the Universe will end their
evolution as white-dwarf stars.  Applications of white dwarfs include
cosmochronology, evolution of planetary systems, and also as
laboratories to study non-standard physics and crystallization. In
addition to the knowledge of their surface properties from
spectroscopy combined with  model atmospheres, the global pulsations
that they exhibit during several phases of their evolution allow
spying on the deep interior of these stars.  Indeed, by means of
asteroseismology, an approach based on the comparison between the
observed pulsation periods of variable white dwarfs and the periods
predicted by representative theoretical models, we can infer details
of the internal chemical stratification, the total mass, and even the
stellar rotation profile and strength of magnetic fields.  In this
article, we review the current state of the area, emphasizing the
latest findings provided by space-mission data.
\keywords{stars: oscillations, stars: evolution, white dwarfs}
\end{abstract}

\firstsection 
\section{Introduction}

White-dwarf stars constitute the end fate of $\sim 97$ \%  of all
the stars that populate the Universe, included our Sun.
Indeed, all the stars with a mass lower than $\sim 8 M_{\odot}$
in the Main Sequence (MS) will passively end their lives as white dwarfs.
The reader interested in details about the formation and evolution of
white dwarfs can consult the review article by \cite{althaus2010}.
Here, we will only give a brief review of the main characteristics of these
stars. White dwarfs are compact stars, characterized by mean densities 
of the order of $\overline{\rho} \sim 10^6$ gr/cm$^3$
(in comparison, $\overline{\rho_{\odot}}= 1.41$ gr/cm$^3$) and radii 
of roughly  $R_{\star} \sim 0.01 R_{\odot}$. They cover a wide interval
of surface temperatures ($4000 \lesssim T_{\rm eff} \lesssim 200\,000$ K)
and, consequently, a vast range of luminosities
($0.0001 \lesssim L_{\star}/L_{\odot}\lesssim 1000$). The mass distribution 
of white dwarfs is characterized by a very pronounced peak at 
$M_{\star} \sim 0.6 M_{\odot}$ (the average mass), although the mass range
is quite wide ($0.15 \lesssim M_{\star}/M_{\odot} \lesssim 1.3$).
White dwarfs with masses in the range
$0.45 \lesssim M_{\star}/M_{\odot} \lesssim 1.0$ likely have cores made of
$^{12}$C and $^{16}$O, while the less massive ones ($M_{\star}/M_{\odot}
\lesssim 0.45$) probably harbor $^4$He cores, and the most massive ones
($M_{\star}/M_{\odot} \gtrsim 1.0$) could have cores made of
$^{16}$O, $^{20}$Ne and $^{24}$Mg. Due to the high densities
characterizing white dwarfs, the equation of state that governs
most of the structure of a white dwarf is that corresponding to a
Fermi gas of degenerate electrons, which provide most of the pressure.
In turn, the non-degenerate ions contribute to the mass of the star
and the heat content accumulated during the previous evolutionary
stages. Loosely speaking, the evolution of a white dwarf consists in a
gradual cooling, during which the energy sources of nuclear reactions  
are irrelevant, although there are some exceptions, such as very hot
pre-white dwarfs, extremely low-mass (ELM) white dwarfs, and average-mass
white dwarfs coming from low-metallicity progenitors.

\begin{figure}
\centering
\includegraphics[width=0.9\textwidth]{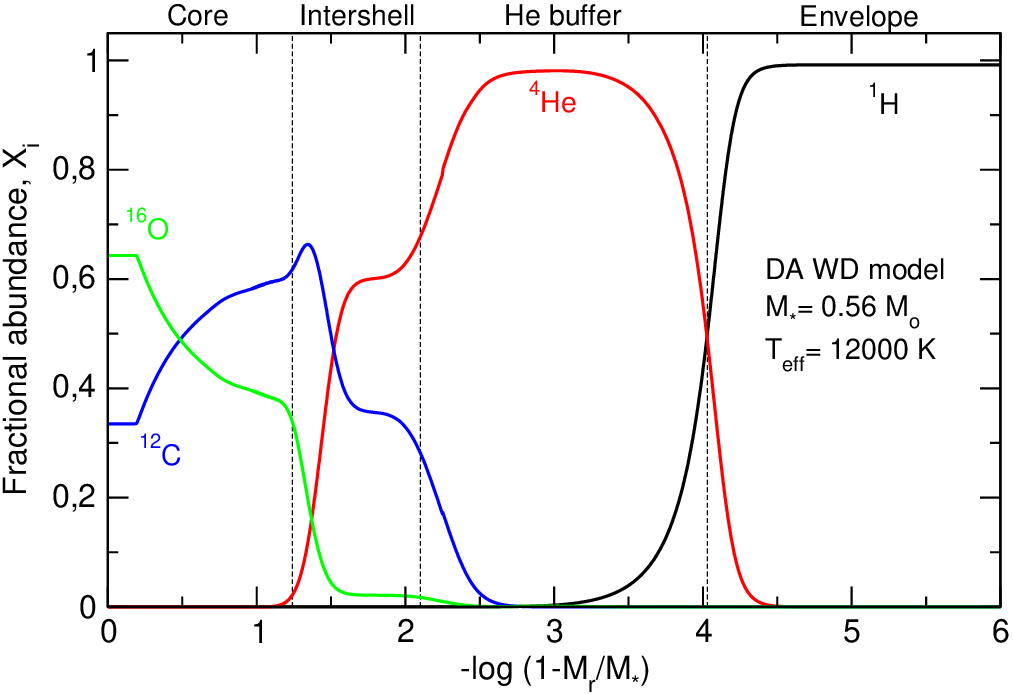}
\caption{Internal chemical structure of a typical DA white dwarf model
  with $M_{\star}= 0.56M_{\odot}$, $T_{\rm eff} \sim 12\,000$ K,
  and H envelope thickness of $\log(M_{\rm H}/M_{\star}) \sim -4$,
  resulting from the complete evolution of a single $1 M_{\odot}$
  progenitor from the ZAMS to the white dwarf stage. Plotted is the mass
  fraction ($X_{\rm i}$) of $^{16}$O
  (green), $^{12}$C (blue), $^{4}$He (red), and $^{1}$H (black),
  in terms of the outer mass fraction coordinate. The centre of the star is at
  $-\log(1-M_r/M_{\star})= 0$. We separate with vertical dashed lines four different parts of the
  internal chemical structure of the model, the origin and the uncertainties
  of which are described in the text.}
\label{fig:01}
\end{figure}

By virtue 
of the high surface gravities ($\log g \sim 8$ [cgs]),
the different chemical species in these stars are well separated due to
the effect of gravitational settling. According to the dominant
chemical species on the surface of white dwarfs, they are classified
into two main groups, those of the spectral type DA ($\sim 80 \%$
of the total, atmospheres rich in H) and those of spectral type DB
($\sim 15 \% $ of the total, atmospheres rich in He).
Figure \ref{fig:01} shows the stratified chemical structure of a DA
white dwarf model as a function of the outer mass coordinate.
This coordinate strongly amplifies the outer regions of the star.
As can be seen, $ 99 \% $ of the total mass of
the star [$-\log (1-M_r / M_ {\star}) \lesssim 2$] is composed mostly
of C and O, the mass of the He layer being 
$M_{\rm He} \lesssim 0.01 M_ {\star} $ and the mass of the H envelope being
 $M_{\rm H} \lesssim 0.0001 M_{\star} $. For illustrative purposes, we
have divided the chemical structure of the star into four parts: the core,
the intershell region, the He-buffer, and the H envelope.
In Table \ref{tab:01}, we include a short account of the origin and the
uncertainties playing a role at each part of the chemical structure
\citep[see, for details,][]{corsico2019}.

\begin{table*}[!ht]
\centering
\caption{The different parts of the chemical structure (first row), their origin (second row),
  and their uncertainties (third row), corresponding to the DA white dwarf model shown
  in Fig. \ref{fig:01}.}
\begin{tabular}{l|c|c|c|c}
\hline
 & Core & Intershell & He buffer & Envelope \\ 
\hline
Origin & The result of He-core     & Built up during & Shaped by               & Shaped by \\
       & burning and the following & the TP-AGB      & gravitational settling. & gravitational settling. \\
       & steady He burning         & phase           & Prior H burning         & Primordial H\\
\hline
Uncertainties & $^{12}$C$(\alpha,\gamma)^{16}$O reaction & Number of TPs, & Number of TPs,  & Metallicity,  \\
              & rate,                                  & extra-mixing   & IFM             & C dredge-up,  \\
              & extra-mixing                           & rotation (?)   & relationship    & late He flashes  \\
\hline
\end{tabular}
\label{tab:01}
\end{table*}

Since white dwarfs are very old objects, with ages of the order of
$\tau \sim 10^9 - 10^{10}$ years, they convey extremely valuable
information about the evolution of stars ``from the cradle to the
grave'', and about the rate of stellar formation along the complete
history of our Galaxy. The study of white dwarfs has many
applications to several fields. One one hand, they can provide an
estimate of the age of stellar populations like open and globular
clusters, and the halo and the galactic disk, by using the luminosity
function of white dwarfs \citep[see, for instance,][]{garcia2010}.
On the other hand, due to the existence of a mass limit
for white dwarfs (the Chandrasekhar mass), they are of utmost
relevance as progenitors of supernovae of type Ia, and also
cataclysmic variables like novae, that involve very energetic events
($E \sim 10^{44} - 10^{51}$ erg) of mass transfer onto the white dwarf
surface from its companion star.  Finally, by virtue of the extreme
conditions of pressure and density prevailing inside white dwarfs,
they allow the investigation of phenomena such as crystallization due
to Coulomb interactions, the assessment of constraints on fundamental
particles like axions and neutrinos, the variation of fundamental
constants, etc. Classically, the approaches employed to study white
dwarfs have been spectroscopy, photometry and astrometry, that allow
us to infer the effective temperature, surface gravity and composition,
parallaxes and distances, etc. Of particular interest in this article
is the technique named ``asteroseismology'', that allows us to extract
information of the {\it interior} of stars by studying how they
pulsate. In the next section we provide a brief description of this
technique applied to pulsating white-dwarf stars.

\section{Stellar pulsations and white-dwarf asteroseismology}
\label{stellar_pulsations}

Asteroseismology is based on the well-known physical principle
establishing that \emph{``studying how a system vibrates in its normal
  modes allows us to infer its mechanical properties''}. In the case of
stars, the vibrations are global pulsations   that allow us to ``see''
the sub-photosferic layers, otherwise inaccessible by means  of
classical techniques\footnote{The only alternative way to obtain
  direct information from the stellar interior is through neutrinos,
  which escape from the depths of the stars without interacting with
  matter.}.  Basically, asteroseismology consists in the comparison of
theoretical periods of stellar models with the periods observed in
real pulsating stars. In the case of white-dwarf asteroseismology,
there are other tools to extract information for several quantities
---in particular the stellar mass--- on the basis of the asymptotic
behavior of $g$ (gravity) modes of high radial order. These tools will
be described in Section \ref{tools}. White-dwarf asteroseismology
allows us to derive information about the stellar mass, the chemical
stratification, the core chemical composition, the existence and
strength of magnetic fields, the properties pf stellar rotation, the
physics of convection, etc. Very detailed reports about white-dwarf
asteroseismology can be found in the review articles by
\cite{fon2008,win2008,althaus2010} and \cite{corsico2019}.

\subsection{A little bit on nonradial stellar pulsations}
\label{nonradial_pulsations}

Stellar pulsations are eigenmodes of the stars that can be thought of
as standing waves in 3 dimensions. Each star has a unique spectrum of
discrete eigenfrequencies, the natural frequencies of the star, that
is set by its internal structure, mass, effective temperature, radius,
etc. In a sense, these natural frequencies are like the
``fingerprints'' of the star.  The eigenfrequencies are associated with
eigenfunctions that provide the spatial variation of the different
physical parameters of the star when it pulsates.  In white-dwarf
stars, the frequency spectrum of pulsations is extremely sensitive to
the details of the internal chemical stratification. Pulsations in
stars can be radial, which retain the spherical symmetry (Cepheids, RR
Lyrae, Miras, etc). Radial pulsations are a particular case of a very
general class of oscillatory movements, called nonradial
pulsations. The latter do not preserve the spherical
symmetry. Nonradial pulsations are routinely detected in the Sun, and
also in variable solar-type stars, red giants, $\delta$ Scuti,
$\gamma$ Doradus, $\beta$ Cephei, SPB, WR, sdB, white dwarfs and
pre-white dwarf stars.  The detection of nonradial pulsations has
experienced an unprecedented flourishing in the last years thanks to
the  advent of the space missions, such as  MOST \citep{2003PASP..115.1023W},
CoRoT \citep{2009IAUS..253...71B}, {\it Kepler}
\citep{2010Sci...327..977B} and K2 \citep{2014PASP..126..398H}
---which have already ended--- the TESS (Transiting Exoplanet Survey
Satellite) mission  \citep{2015JATIS...1a4003R} ---that is currently working---
and the future missions Cheops \citep{2018A&A...620A.203M} and
Plato \citep{2018EPSC...12..969P}, among others.

In the frame of the linear theory, that assumes small amplitudes of
pulsation, the deformations of a star when it pulsates in
spheroidal modes are specified by the Lagrangian displacement vector
\citep{unno1989}:

\begin{equation} 
\vec{\xi}_{k \ell m}= \left(\vec{\xi}_r, \vec{\xi}_{\theta}, 
\vec{\xi}_{\phi} \right)_{k \ell m}
\end{equation}

\noindent where the components in spherical coordinates are:

\begin{eqnarray}
\vec{\xi}_r & = & \xi_r(r) Y^m_{\ell}(\theta, \phi) e^{i \sigma t} \vec{e}_r \\
\vec{\xi}_{\theta} & = & \xi_h(r) \frac{\partial Y^m_{\ell}}{\partial \theta} e^{i \sigma t} \vec{e}_{\theta} \\
\vec{\xi}_{\phi} & = & \xi_h(r) \frac{1}{\sin \theta} \frac{\partial Y^m_{\ell}}{\partial \phi} e^{i \sigma t} \vec{e}_{\phi}  \\
\nonumber
\end{eqnarray}

\noindent Here, $Y^m_{\ell}(\theta, \phi)$ are the spherical
harmonics,  $\sigma$ is the pulsation frequency, and  $\xi_r(r)$ y
$\xi_h(r)$ are the radial and horizontal eigenfunctions,
respectively. Each eigenmode has a sinusoidal temporal dependence, an
angular dependence by means of spherical harmonics, and a radial
dependence given through the radial and horizontal eigenfunctions,
which must inevitably be obtained (for realistic star models) through
the numerical resolution of the differential equations of stellar
pulsations \citep[see][]{unno1989}. Pulsation  modes are characterized
by three ``quantum numbers'': {\it (i)} harmonic degree $\ell = 0, 1,
2, 3, \cdots, \infty$, that represents $(\ell-m)$ nodal lines
(parallels) on the stellar surface, {\it (ii)} azimuthal order $m=
-\ell, \cdots, -2, -1, 0 , +1, +2, \cdots, +\ell$, that represents $m$
nodal lines (meridians) on the stellar surface, and {\it (iii)} the
radial order $k= 0, 1, 2, 3, \cdots, \infty$, that represents
concentric spherical nodal surfaces on which the fluid displacement is
null. Within the set of spheroidal modes\footnote{Spheroidal modes are
  characterized by $(\vec{\nabla} \times \vec{\xi})_r= 0$ and
  $\sigma^2= 0$, where $\vec{\xi}$ is the Lagrangian displacement
  vector and $\sigma$ the pulsation frequency \citep{unno1989}.},
there are two families of modes that differ according to the dominant
restoring force. The $p$ (pressure) modes, on one hand, involve large
variations of pressure and displacements mostly  radial,
compressibility being the dominant restoring force. They  are
characterized by short periods (high frequencies). The $g$ (gravity)
modes, on the other hand, are associated with small variations of
pressure and large tangential displacements. In this
case, the restoring force is gravity through buoyancy, and the modes
are characterized by long periods (small frequencies).  These are the
modes usually detected in white dwarfs. In the case in which $\ell >
1$, there exist a third sub-class of modes, the $f$ modes. They have
intermediate characteristics between those of the $p$ and $g$ modes,
and  do not have radial nodes ($k= 0$), except for stellar models with
very high central densities.

\subsection{White-dwarf pulsations}
\label{wd_pulsations}

\subsubsection{A brief historical account}

The first theoretical hints of pulsations in white dwarfs were revealed by the pioneering work of \cite{sauvenier1949} and \cite{ledoux1950}, who found instability in
radial ($\ell= 0$) modes with periods of $\sim 10$ s  due to hydrogen nuclear burning. 
These periods were not detected at that time, and this led to the conclusion that H 
burning is not the source of energy in a white dwarf, and allowed \cite{mestel1952} to
elaborate his theory of cooling. The first pulsating white dwarf, HL Tau 76,
was discovered by \cite{landolt1968}, with a detected period of $\sim 740$ s,
too long as to be due to a radial mode \citep{faulkner1968,ostriker1968}.
Two additional pulsating white dwarfs, G44$-$32 (600-800 s) and
R548 (200-300 s) \citep{lasker1969,lasker1971} were discovered shortly after, but again,
the periods measured were too long as to be associated with radial modes. The first
demonstration that the periods detected in these white dwarfs were due to
nonradial $g$ modes
came from \cite{warner1972}  and \cite{chanmugam1972}, in agreement with 
the theoretical work of \cite{baglin1969}  and \cite{harper1970}. An additional
hint that these periods were due to nonradial modes came from the detection of rotational
splitting in the observed frequencies \citep{robinson1976}, since
radial-mode frequencies do not exhibit splitting due to rotation. A definitive proof of
the nonradial $g$-mode nature of the periods of pulsating white dwarfs was provided by
the important work of \cite{mcgraw1979}, who confirmed that the variability is due to changes in
surface temperature and not to changes of the radius, which is typical of $g$ modes. Finally, 
\cite{robinson1982} demonstrated that the variations in the stellar radius
are quite small ($\Delta R_{\star} \sim 10^{-5} R_{\star}$), and that the changes in effective
temperature ($\Delta T_{\rm eff} \sim 200$ K) is what matters most in the pulsations of
white dwarfs. 

\begin{table*}[!ht]
\caption{Properties of the different sub-classes of variable pulsating white dwarfs and pre-white dwarfs,
  sorted by decreasing effective temperature. The tentative classes of pulsators are labeled with a
  question mark in parentheses.}
\begin{tabular}{llccccc}
\hline
\noalign{\smallskip}
Class    &   Year         &  $T_{\rm eff}$         & $\log g$ &  Periods & Amplitudes &  Main surface  \\
         &  of disc. (\#) &   [$\times$ 1000 K]  & [C.G.S.] & [s]      &   [mag]      & composition\\
\noalign{\smallskip}
\hline
\noalign{\smallskip}
GW Vir (PNNV) & 1984 (10)  & $100-180$  & $5.5-7$    & $420-6000$ & $0.01-0.15$ & He, C, O \\
GW Vir (DOV)  & 1979 (9)   & $80-100$   & $7.3-7.7$  & $300-2600$ & $0.02-0.1$  & He, C, O \\
              &            &            &            &            &             &          \\
Hot DAV (?)      & 2013 (3)   & $30-32.6$  & $7.3-7.8$  & $160-705$   & $0.001-0.015$ & H   \\ 
&&&&&&\\
V777 Her (DBV) & 1982 (27) & $22.4-32$  & $7.5-8.3$  &  $120-1080$  & $0.05-0.3$  & He (H) \\
&&&&&&\\
DQV (?)       &  2008 (6)  & $19-22$    & $8-9$  & $240-1100$  & $0.005-0.015$& He, C   \\
&&&&&&\\
GW Lib & 1998 (20) & $10.5-16$ & $8.35-8.7$ & $100-1900$  & $0.007-0.07$ &  H, He \\
&&&&&&\\
ZZ Cet (DAV) & 1968 (260) &  $10.4-12.4$  & $7.5-9.1$  & $100-1400$ &  $0.01-0.3$  & H     \\
&&&&&&\\
pre-ELMV      & 2013 (5)   & $8-13$     & $4-5$  &  $300-5000$      & $0.001-0.05$ & He, H   \\
&&&&&&\\
ELMV          & 2012 (11)  & $7.8-10$   & $6-6.8$  & $100-6300$       & $0.002-0.044$  &  H    \\
\noalign{\smallskip}
\hline
\end{tabular}
\label{table:02}
\end{table*}

\subsubsection{The present situation}

\begin{figure}
\includegraphics[width=0.90\textwidth]{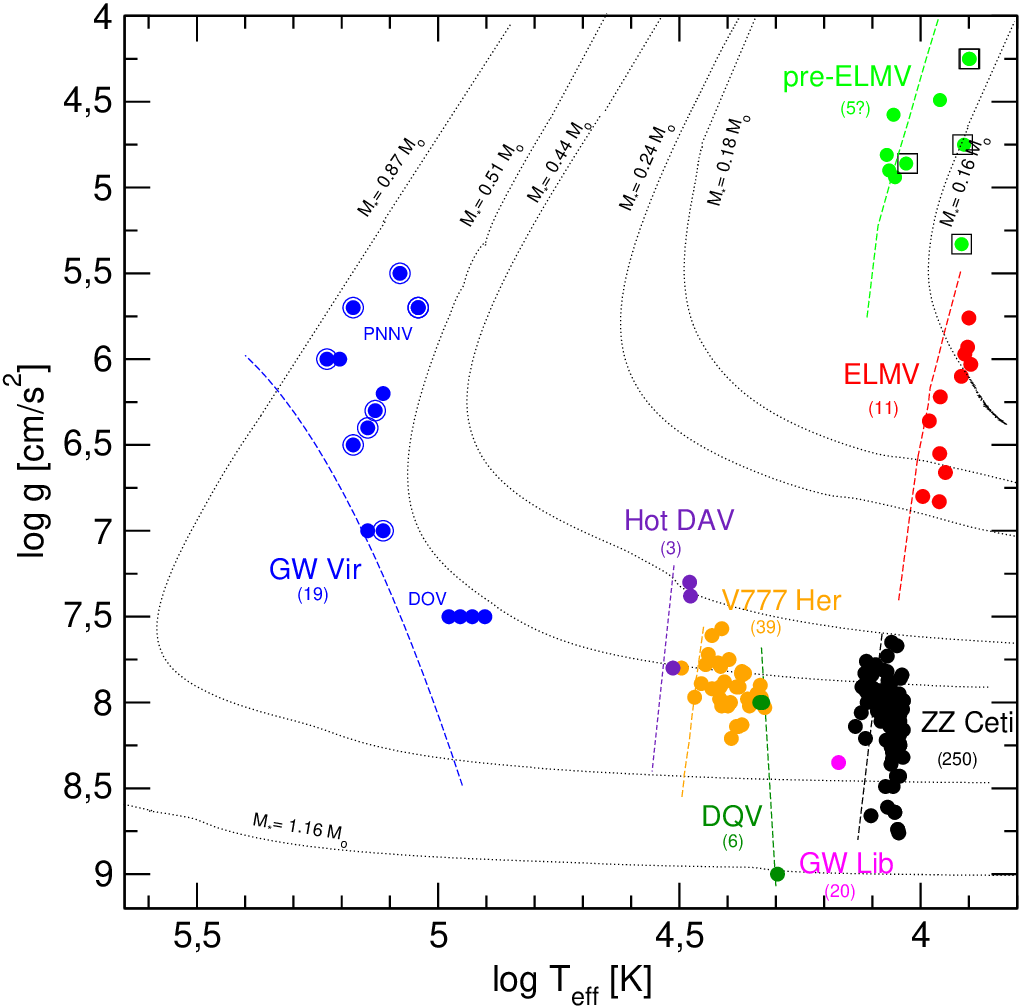}
\caption{Confirmed and tentative sub-classes of pulsating  white-dwarf and
  pre-white-dwarf stars (circles of different colors) in the $\log T_{\rm eff}-
  \log g$ diagram \citep[adapted from][]{corsico2019}. Stars emphasized with squares
  surrounding the light  green circles can be identified as pre-ELMV
  stars as well as SX Phe  and/or $\delta$ Scuti stars. GW Vir stars
  indicated with blue circles surrounded by  blue circumferences are
  PNNVs.  In the case of GW Lib stars, only the location of the prototypical 
  object, GW Librae, has been included (magenta dot).
  Two post-VLTP (Very Late Thermal Pulse) evolutionary tracks
  for H-deficient white dwarfs  \citep[$0.51$ and $0.87 M_{\odot}$;][]{miller2006},
  four evolutionary tracks  of
  low-mass He-core H-rich white dwarfs \citep[$0.16$, $0.18$, $0.24$, and
    $0.44 M_{\odot}$;][]{althaus2013}, and one evolutionary
  track for  ultra-massive H-rich white dwarfs \citep{camisassa2019}
  are  plotted for   reference. Dashed lines indicate the theoretical blue edge of the
  instability domains.}
\label{fig:02}
\end{figure}

Currently, it is a well established
fact that, along their evolution, white dwarfs go through at least one stage of pulsation
instability in which they become pulsating variable stars, showing light curves with
variations in the optical and in the far UV parts of the electromagnetic spectrum
\citep{win2008,fon2008,althaus2010,corsico2019}.
The changes in brightness,
with amplitudes between 0.001 mmag and 0.4 magnitudes, are due to $g$ modes with harmonic
degree $\ell= 1, 2$. In general terms, $g$ modes in white dwarfs probe the envelope
regions, due to the very low values of the Brunt-V\"ais\"al\"a frequency (the critical
characteristic frequency for the  $g$-mode spectrum) in the core regions. This is 
contrary to what happens in normal (non-degenerate) pulsating stars.
At present, more than 350 pulsating white dwarfs and pre-white dwarfs
have been discovered through observations from Earth, in their great majority extracted
from the Sloan Digital Sky Survey (SDSS), and in the last  years also through space
missions such as the already finished {\it Kepler}/K2 Mission, and currently by TESS
\citep{corsico2019}. Pulsating white dwarfs  exhibit a wide variety of light curves,
some sinusoidal and with small amplitudes,
others nonlinear and with large amplitudes. They are multimodal pulsating stars
(they pulsate in more than one period), and frequently exhibit harmonics and linear
combinations of eigenfrequencies that are not related to genuine modes of pulsation, but
instead to nonlinear effects. At present, there are six classes of confirmed pulsating
white dwarfs and pre-white dwarfs known (ZZ Ceti or DAV stars, GW Lib stars, V777 Her or DBV
stars, GW Vir stars, ELMV stars and pre-ELMV stars) and two tentative classes of
white-dwarf pulsators, that need confirmation (hot DAV stars and DQV stars). In Table
\ref{table:02} we show in a compact way the main characteristics of these families of
pulsating white dwarfs and pre-white dwarfs, while in Figure \ref{fig:02} we show their
location in the $\log T_{\rm eff} - \log g$ diagram \citep[details can be found in][]{corsico2019}. In Table \ref{table:02}, the second column indicates the discovery year of the first
object of each class and number of known objects at the date of
writing this article (November 2019), the third column shows the range of effective temperatures
at which they are detected (instability strip), the fourth column provides the range of surface gravity,
the fifth column indicates the range of periods detected, the sixth column contains the range of
amplitudes of the variations in the light curves, and the seventh column shows the surface composition.
Pulsation periods are usually between $\sim 100$ s and $\sim 1400 $s, although PNNVs and ELMVs exhibit
much longer periods, up to $\sim 6300 $ s. Interestingly, the periods of $g$ modes of
white dwarfs are of the same order of magnitude as the periods of $p$ modes in non-degenerate
pulsating stars.

\subsection{The tools of white-dwarf asteroseismology}
\label{tools}

The tools employed in asteroseismology of pulsating white dwarf stars have been
described in several review articles (Winget \& Kepler 2008, Fontaine \& Brassard 2008, Althaus et al.
2010). Here, we will focus on the information that the existence of a constant (or nearly constant)  
period separation among the periods observed for a given star can give about the stellar mass, in particular
in the case of GW Vir stars and DBVs. Also, we will describe the search for asteroseismological models
that replicate the individual observed periods in real stars, that is, the period-to-period fits. Finally,
we will revise the information about the stellar rotation that is possible to extract from the frequency splittings.

\subsubsection{Clues to the stellar mass from the period spacing}
\label{per_spa}

\begin{figure}  
\centering  
\includegraphics[clip,width=350pt]{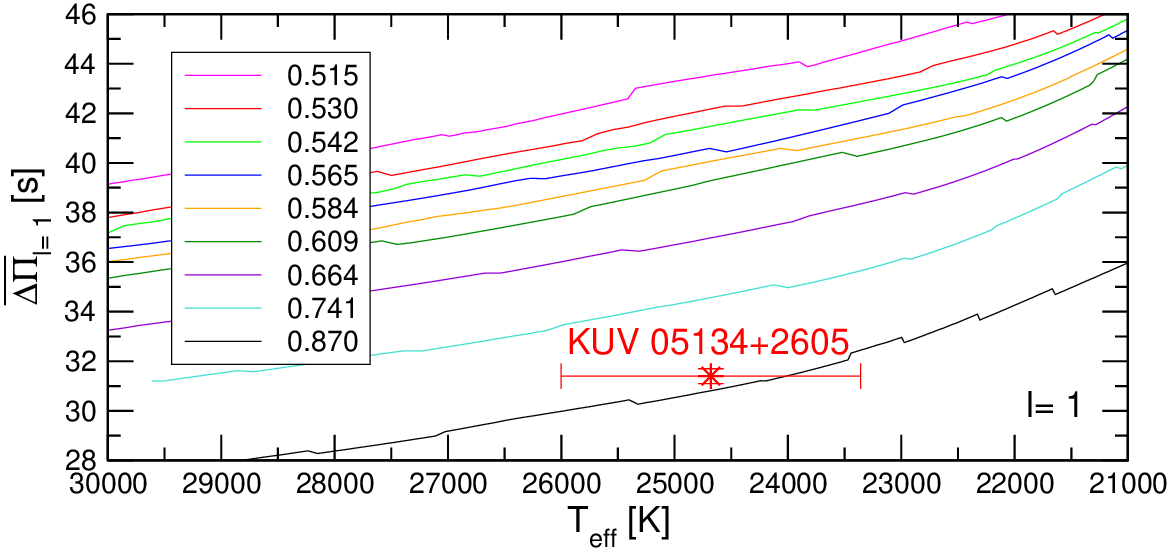}  
\caption{Average of the  computed period spacings for $\ell= 1$ 
corresponding to DB white  dwarf sequences  with different stellar 
masses. The location of the target star  KUV 05134+2605  is shown with  a red star  
symbol ($T_{\rm eff} = 24\,700 \pm 1300$ K) for each solution of observed 
mean period spacing, $\Delta \Pi_{\ell= 1}= 31.4 \pm 0.3$ s. By simple linear interpolation, we 
found that the mass of the star according to its period spacing is 
$M_{\star}= 0.85\pm 0.05M_{\odot}$.}
\label{fig:03}  
\end{figure}

For  $g$-modes  with  high radial  order  $k$  (long  periods),  the separation  of consecutive
periods ($|\Delta  k|= 1$)  becomes nearly constant  at a  value  given  by the  asymptotic
theory of  nonradial stellar  pulsations.   Specifically,  the  asymptotic  period  spacing
\citep{1990ApJS...72..335T} is given by:

\begin{equation} 
\Delta \Pi_{\ell}^{\rm a}= \Pi_0 / \sqrt{\ell(\ell+1)},  
\label{aps}
\end{equation}

\noindent where

\begin{equation}
\label{asympeq}
\Pi_0= 2 \pi^2 \left[ \int_{r_1}^{r_2} \frac{N}{r} dr\right]^{-1},
\end{equation}

\noindent $N$ being the Brunt-V\"ais\"al\"a frequency. 
This expression is rigorously valid for chemically homogeneous
stars.  In  principle, one can  compare the asymptotic  period spacing
computed from  a grid  of models with  different masses  and effective
temperatures with the  mean period spacing exhibited by  the star, and
then infer the value of the stellar mass. This method has been applied
in numerous  studies of  pulsating PG 1159  stars
\citep[see,  for instance,][and references therein]{corsico2007a,corsico2007b,
  corsico2008,corsico2009}. For the
method to be  valid, the periods exhibited by  the pulsating star must
be associated  with high  order $g$ modes, that  is, the star  must be
within  the asymptotic  regime  of pulsations.   
Furthermore, the interior  of white-dwarf  stars are
supposed  to  be chemically  stratified  and  characterized by  strong
chemical gradients built up during  the progenitor star life.  So, the
direct  application of  the  asymptotic period  spacing  to infer  the
stellar  mass of  white dwarfs is  somewhat  questionable.   A better way
to compare the models to the period spacing  of the observed  pulsation spectrum,
is to  calculate the average of the computed period spacings, using:

\begin{equation}
\label{avgdp}
\overline{\Delta  \Pi}(M_{\star}, T_{\rm eff})= \frac{1}{(n-1)} 
\sum_{k}^{n-1}  \Delta \Pi_k, 
\end{equation}

\noindent  where  $\Delta \Pi_k$  is  the  ``forward'' period  spacing
defined as $\Delta  \Pi_k= \Pi_{k+1}-\Pi_k$, and $n$ is  the number of
computed periods  laying in  the range of  the observed  periods.  The
theoretical period spacing  of the models as computed 
through Eq.  (\ref{aps}) and (\ref{avgdp})
share the same  general trends (that is, the  same dependence on $M_{\star}$
and $T_{\rm  eff}$), although  $\Delta \Pi_{\ell}^{\rm a}$  is usually
somewhat higher  than $\overline{\Delta \Pi}$, particularly for the case of
low radial-order $g$ modes. We can compare the measured mean period spacing of
a given pulsating white dwarf with the average of the computed period spacings
for several sequences with different stellar masses, by fixing the effective
temperature and its uncertainties for the target star
derived from spectroscopy. In this way, an estimate of the stellar mass
can be obtained. As an example of application, in Fig. \ref{fig:03} we show the
run of the average of the computed period spacings [Eq. (\ref{avgdp})] with $\ell= 1$, 
in terms  of the effective temperature for a set of DB white dwarf 
evolutionary  sequences.
In the plot, we have included the measured period
spacing for a DBV star, KUV 05134+2605 \citep[$T_{\rm eff}= 24\,700\pm1300$ K and
$\log g= 8.21\pm0.06$;][]{2011ApJ...737...28B},
whose period spectrum exhibits $\Delta \Pi_{\ell= 1}= 31.4 \pm 0.3$ s. 
By means of a linear interpolation, we obtain the mass of the star according
to its period spacing, $M_{\star}= 0.85\pm 0.05M_{\odot}$.

A cautionary word is necessary here. In the case of GW Vir stars and DBV stars,
the period spacing is sensitive mostly to the stellar mass and effective temperature
and very weakly to the thickness of the He envelope in the case of DBVs
\citep[see][]{1990ApJS...72..335T} and the thickness of the C/O/He envelope in the case of
the GW Vir stars \citep{1994ApJ...427..415K}. In the case of ZZ Ceti stars, however, the
period spacing depends on the stellar mass, the effective temperature, and the thickness of
the H envelope with a comparable sensitivity. Consequently, the method is not
---in principle--- directly
applicable to ZZ Ceti stars due to the intrinsic degeneracy of the dependence of $\Delta \Pi$ with the
three parameters $M_{\star}$, $T_{\rm eff}$, and $M_{\rm H}$ \citep{fon2008}.
An illustration
of this has been recently given in the asteroseismological study of ultra-massive ZZ Ceti stars 
by \cite{2019arXiv191007385C}.

\subsubsection{Constraints from period-to-period fits}
\label{per_to_per}

\begin{figure}  
\centering  
\includegraphics[clip,width=350pt]{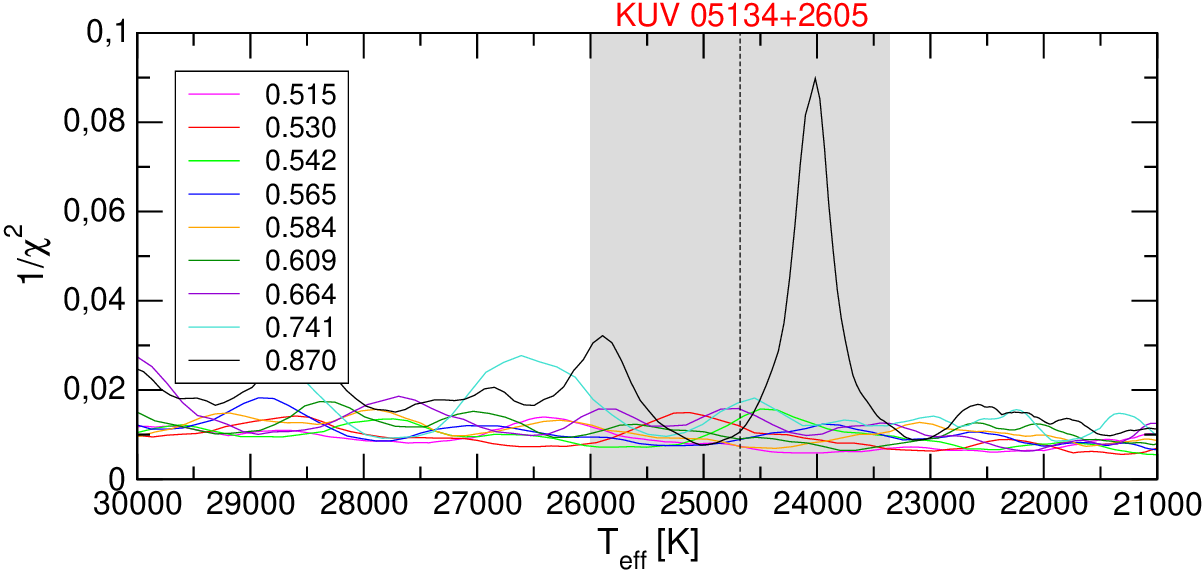}  
\caption{Inverse of the quality function of the period 
fit in terms of the effective temperature. The vertical gray strip 
indicates the spectroscopic $T_{\rm eff}$ of KUV 05134+2605 and its uncertainties 
($T_{\rm eff} = 24\,680 \pm 1322$ K). The strong maximum
in $(\chi^2)^{-1}$ corresponds to the best-fit model, with
$M_{\star}= 0.870  M_{\odot}$ and $T_{\rm eff}= 23\,  976$ K.}
\label{fig:04}  
\end{figure}  

Another  way to  infer the  stellar mass,  along with the 
effective  temperature and also many details  of the  internal  structure  
of pulsating white dwarfs, is through  their individual pulsation  periods.   
In  this approach, we  seek a pulsation white dwarf model  
that best matches the  pulsation periods  of the target star. The  
goodness of  the  match between  the
theoretical  pulsation periods ($\Pi_k$)  and the  observed individual
periods  ($\Pi_{{\rm obs},  i}$) is  measured  by means  of a  quality
function defined as: 

\begin{equation}
\chi^2(M_{\star}, T_{\rm eff})= \frac{1}{N} \sum_{i=1}^{N} 
\min[(\Pi_{{\rm obs},i}- \Pi_k)^2], 
\label{ptpf}
\end{equation}

\noindent where $N$   is the number of  observed periods.  The 
white dwarf model  that shows the lowest value  of $\chi^2$ is adopted
as the ``best-fit model'' \citep[see][]{corsico2007a,corsico2007b,corsico2008,
  corsico2009,corsico2012,2019A&A...632A..42B}. The quality of the period fits is
assessed by
means   of   the  average   of   the   absolute  period   differences,
$\overline{\delta}=  (\sum_{i=1}^N  |\delta_i|)/N$,  where  $\delta_i=
\Pi_{{\rm  obs}, i}  -\Pi_k$,  and by  the root-mean-square  residual,
$\sigma= \sqrt{(\sum |\delta_i|^2)/N}= \sqrt{\chi^2}$. Also, in
order  to have an indicator of the quality of  the 
period fit,  we can compute the  Bayes Information
Criterion \citep[BIC;][]{2000MNRAS.311..636K}:

\begin{equation}
{\rm BIC}= N_{\rm p} \left(\frac{\log N}{N} \right) + \log \sigma^2,
\end{equation}

\noindent where $N_{\rm p}$ is  the number of free parameters, and $N$
the number of  observed periods. The  smaller the value of  
BIC, the better the quality  of the fit. Following the same example 
of the previous section, in Fig. \ref{fig:04} we show
the inverse of the quality function of the period fit for the DBV star
KUV 05134+2605 in terms of the effective temperature. The vertical gray strip indicates
the spectroscopic $T_{\rm eff}$ and its uncertainties. Clearly,
the model that best reproduces the observed periods has a stellar mass
of $0.87 M_{\odot}$ and an effective temperature $T_{\rm eff}\sim 24\,000$ K.

\subsubsection{Rotational splittings}

When a pulsating star is rotating, the frequencies of nonradial
pulsation modes split into $2\ell+1$ components, and their separation
is proportional to the velocity of rotation of the star. In this way,
when a pulsating white dwarf exhibits rotational splitting, it is
possible to constrain its rotation period. If the rotation is slow and
rigid, the rotation frequency $\Omega$ of the white dwarf is connected
with the frequency  splitting $\delta \nu$ through the coefficients
$C_{k, \ell}$ ---that depend on the details of the stellar
structure--- and the values of $m$ ($-\ell, \cdots, -1, 0, +1, \cdots,
+\ell$), by means of the expression $\delta \nu = m
(1-C_{k,\ell})\ \Omega$ \citep{unno1989}. Note that, when present,
frequency splitting ($\ell= 1$: triplets, $\ell= 2$: quintuplets, etc)
allows us to identify the harmonic degree $\ell$ and the azimuthal order
$m$ of the modes.  Thanks to this effect, the rotation rate of many
pulsating white dwarf stars has been measured (see next Section).

\section{Recent findings}
\label{findings}

In this section, we describe the most recent results of white-dwarf
asteroseismology. In particular, we will
focus on outstanding findings that have been the result of
uninterrupted observations from space, in particular, through the {\it
  Kepler}/K2 Mission. These findings would not have been possible
through observations from the ground. We mainly describe the results
from several relevant works that study the pulsation properties of
V777 Her and ZZ Ceti stars from space-based observations. We also
describe the first asteroseismological analysis carried out on the DBV
star EC01585$-$1600 with the TESS
data. Finally, we briefly report on a  investigation of the
possible existence of pulsating warm ($T_{\rm eff} \sim 19\,000$ K) DA
white dwarfs, and the null results from TESS. 

\subsection{Outbursting ZZ Ceti stars}

The {\it Kepler} spacecraft observations of ZZ Ceti stars revealed a
new type of phenomenon never observed before from the ground:
outburst-like events in DAV stars \citep{2017ASPC..509..303B}. The first one,
WD J1916+3938 (KIC4552982), was observed during 1.5 years. 20
pulsation modes with periods typical of ZZ Ceti stars were detected,
along with 178 enhancements of brightness typical of outburst
phenomena, with peaks of up to 17 \% above the quiescent level, 
involving very energetic events ($E\sim 10^{33}$ erg), with a mean
recurrence period of about 2.7 days \citep{2015ApJ...809...14B}.  The second
outbursting ZZ Ceti star, PG1149+057, was studied by \cite{2015ApJ...810L...5H}.
This star shows flux enhancements of up to 45\% above the
quiescent level! It was revealed that for this star, the outbursts
actually affect the normal pulsations (in amplitude and
frequency). This  leads to the important conclusion that outbursts
are an intrinsic phenomenon of the (otherwise isolated) star. At present,
a total of 8 outbursting isolated ZZ Ceti stars have been discovered
\citep{2016ApJ...829...82B,2017ASPC..509..303B}, close to the red edge of the ZZ
Ceti instability strip. A possible explanation is connected with
parametric instability via mode coupling of white dwarf pulsations
\citep{1982AcA....32..147D,2001ApJ...546..469W,2018ApJ...863...82L}.

\subsection{Dichotomy of mode line widths in ZZ Ceti stars}

\cite{2017ApJS..232...23H} discovered a dichotomy of mode line widths in
the power spectrum of pulsating DA white dwarfs observed from
space. Specifically, low-frequency modes with periods longer than
roughly 800 s are generally incoherent over the length of
observations, while higher-frequency modes (< 800 s) are observed to
be much more stable in phase and amplitude. This phenomenon is
incompatible with stochastic excitation, that for white dwarfs would
be able to excite pulsations with periods of the order of $\sim 1$
sec. 27 DAVs were observed through K2 Campaign by the {\it Kepler}
space telescope in the study of \cite{2017ApJS..232...23H}.
The dichotomy of model line width can be related to
the oscillation of the outer convection zone of a DA white dwarfs
during pulsations. The oscillation of the base of the outer
convection zone would affect $g$-mode eigenfunctions having the outer
turning point located at the base of the convection zone \citep{2019arXiv190205615M}.
Deepening the study of this phenomenon can lead to a
better understanding of the properties of the outer convection zone of
DA white dwarfs.

\subsection{White-dwarf rotation rates from asteroseismology}

The rotation rates of 27 isolated DA white dwarfs have been determined
via asteroseismology using rotational splittings by \cite{2017ApJS..232...23H}
with observations from {\it Kepler}/K2 Mission.  In this
way, the number of pulsating white dwarfs with measured rotation rates
has been doubled \citep[see Table 10 of][]{corsico2019}.  The results
indicate that the mean rotation period is of $\sim 30-40$ hours,
although fast rotating stars have also been found, with periods of
$\sim 1-2$ hours. Evidence has been found for a link between high
mass and fast rotation, although additional massive white dwarfs are
required to confirm this trend. If true, this trend would favor
the hypothesis that the high-mass white dwarfs are the result of
mergers, which are supposed to rotate very fast. \cite{2017ApJS..232...23H}
find that the majority of isolated descendants of $1.7–3.0 M_{\odot}$
ZAMS progenitors rotate at $\sim 1.5$ days, instead of minutes! This
indicates that most internal angular momentum must be lost on the
first-ascent giant branch.

\subsection{Amplitude and frequency variations of $g$ modes in a DBV white dwarf}

By using observations from the {\it Kepler} mission,  \cite{2016A&A...585A..22Z}
have detected amplitude and frequency variations of the components of
the triplets of frequencies caused by rotation in the DBV star KIC
08626021. The timescale for the quasi-periodic modulations of the
variations is of about 600 days, undetectable from ground-based
observations. A similar phenomenon has been detected with {\it Kepler}
in the pulsating sdB star KIC 10139564  \citep{2016A&A...594A..46Z}.  It is
thought that the modulations of frequencies and amplitudes are not
related to any evolutionary effect (e.g., neutrino cooling), since the
timescales involved are several orders of magnitude shorter than the
cooling rate of DB white dwarfs. These modulations cannot be
attributed to signatures of orbiting companions around the star,
because different timescales for different triplets are
detected. Therefore, a possible explanation  is nonlinear resonant
mode coupling in rotationally split triplets
\citep{1997A&A...321..159B,1998BaltA...7...21G}.
If so, this star could constitute a new window
to study nonlinear pulsations. On the other hand, the detected
frequency modulations can potentially prevent a measurement of the
evolutionary (cooling) rate of period change of the star.

\subsection{The DBV pulsator WD0158$-$160: TESS observations}

The first pulsating white dwarf analyzed with TESS is the DBV pulsator
WD0158$-$160 \citep[also called EC01585$-$1600, G272$-$B2A, TIC257459955;][]
{2019A&A...632A..42B}. TESS performs extensive time-series photometry that
allows to discover pulsating stars, and, in particular, pulsating
white dwarfs. The TESS Asteroseismic Science Consortium (TASC)
Working Group 8 (WG8) focuses on TESS observations of evolved compact
stars that exhibit photometric variability, including hot subdwarfs,
white dwarfs, and pre-white dwarfs with mag $< 16$, with short (120
sec) cadence. TIC257459955 is a known DBV white dwarf with
$T_{\rm eff}= 24\,100$ K and $\log g= 7.88$ \citep{2018ApJ...857...56R}, or
alternatively, $T_{\rm eff}= 25\,500$ K and $\log g= 7.94$ \citep{2007A&A...470.1079V}.
\cite{2019A&A...632A..42B} find  9 independent frequencies suitable
for asteroseismology plus frequency combinations. The periods of
genuine eigenmodes are in the range $[245-866]$ sec, with a period
spacing of $\Delta \Pi= 38.1\pm 1.0$ s associated to $\ell= 1$. The
comparison with the average of the computed period spacings (see
Section \ref{per_spa}) gives an estimate of the stellar mass. Here, it
is assumed that the period spacing is primarily dependent on $T_{\rm
  eff}$ and $M_{\star}$ (and very weakly on $M_{\rm  He}$). This
method indicates a stellar mass of $M_{\star}= 0.621 \pm 0.06
M_{\odot}$, or alternatively, $M_{\star}= 0.658 \pm 0.10 M_{\odot}$
(according to the two different spetroscopic determinations of $T_{\rm
  eff}$), larger than the spectroscopic estimates ($M_{\star}=
0.542-0.557 M_{\odot}$). The star also has been analyzed by performing
period-to-period fits (Section \ref{per_to_per}) by employing a
fully evolutionary model approach (La Plata Group) and a
parametric approach (Texas Group). There exist families of
asteroseismic solutions, but the solution that satisfies the
spectroscopic parameters and the astrometric constraints from Gaia, is
characterized by a DB white dwarf model with $M_{\star} \sim 0.60
M_{\odot}$, $T_{\rm eff} \sim 25\,600$ K, $M_{\rm He} \sim 3 \times
10^{-2} M_{\star}$, $d \sim 67$ pc, and a rotation period of $\sim 7$
or $\sim 14$ hours.  It is worth emphasizing that the
asteroseismological solutions from the La Plata Group and the Texas
Group are similar regarding the location of the chemical transition
zones, which are mainly responsible for setting the period spectrum of
a white dwarf model.

\subsection{The possible existence of pulsating warm DA white dwarfs}

In their theoretical pioneering work, \cite{1982ApJ...252L..65W}
\citep[see also][]{1982PhDT........27W}
predicted the possible existence of warm ($T_{\rm eff}
\sim 19\,000$ K) pulsating DA white-dwarf stars, hotter than ZZ Ceti
stars.  However, to date, no pulsating warm DA white dwarf has been
detected. \cite{2019arXiv191102442A}  re-examined the pulsational
predictions for such white dwarfs on the basis of new full
evolutionary sequences and also analyzed a sample of warm DA white
dwarfs observed by the TESS satellite in order to search for the possible
pulsational signals. \cite{2019arXiv191102442A} computed white-dwarf
evolutionary sequences with very small H content, appropriate for the
study of warm DA white dwarfs, employing a new full-implicit treatment
of time-dependent element diffusion. Also, they computed non-adiabatic
pulsations in the effective temperature range of $30\,000-10\,000$ K,
focusing on $\ell= 1$ $g$ modes with periods in the range $50-1500$ s.
These authors found that extended and smooth He/H transition zones
inhibit the excitation of $g$ modes due to partial ionization of He
below the H envelope, and only in the case that the H/He transition is
assumed to be much more abrupt, do the models exhibit pulsational
instability. In this case, instabilities are found only in white dwarf
models with H envelopes in the range of $-14.5 \lesssim \log(M_{\rm
  H}/M_{\star}) \lesssim -10$ and at effective temperatures higher
than those typical of ZZ Ceti stars, in agreement with the previous
study by \cite{1982ApJ...252L..65W}.  \cite{2019arXiv191102442A} found that none
of the warm DAs observed by the TESS satellite are pulsating. This
study suggests that the non-detection of pulsating warm DA white
dwarfs, if white dwarfs with very thin H envelopes do exist, could be
attributed to the presence of a smooth and extended H/He transition
zone.  This could be considered as an indirect proof that element
diffusion indeed operates in the interior of white dwarfs. 

\section{Conclusions}

From the beginning to the present, white-dwarf asteroseismology has
undergone impressive progress, in recent years by the availability of space
missions that provide unprecedented high-quality data, but also from
ground-based observations, mainly with the spectral observations of
the Sloan Digital Sky Survey \citep[SDSS,][]{2000AJ....120.1579Y}. This progress from
the observational front has been accompanied by the development of new
detailed models of white dwarf stars and novel  implementations of
asteroseismological techniques. The combination of these factors is
leading the white-dwarf asteroseismologists to realize the desire to
know the details of the inner structure and evolutionary origins of
these stars. These studies will soon be driven by new observations from
space  (TESS, PLATO, Cheops). Therefore, the future of this
area is, at the very least, very promising.

\acknowledgements I would warmly thanks the Local Organising Committee
of the IAU Symposium 357, in particular to Professor Martin Barstow,
for support that allowed me to attend this conference.


\begin{thebibliography}{}

\bibitem[Althaus et al. (2010)]{althaus2010} Althaus, L. G., C\'orsico, A. H., Isern, J., Garc\'ia-Berro, E.\ 2010, A\&ARv,18, 471   
\bibitem[Althaus et al. (2013)]{althaus2013} Althaus, L.~G., Miller Bertolami, M.~M., \& C{\'o}rsico, A.~H.\ 2013, A\&A, 557, A19  
\bibitem[Althaus et al.(2019)]{2019arXiv191102442A} Althaus, L.~G., C{\'o}rsico, A.~H., Uzundag, M., et al.\ 2019, arXiv e-prints, arXiv:1911.02442
\bibitem[Baglin (1969)]{baglin1969} Baglin, A.\ 1969, Astrophysical Letters, 3, 119
\bibitem[Baglin et al.(2009)]{2009IAUS..253...71B} Baglin, A., Auvergne, M., Barge, P., et al.\ 2009, Transiting Planets, 71
\bibitem[Bell et al.(2019)]{2019A&A...632A..42B} Bell, K.~J., C{\'o}rsico, A.~H., Bischoff-Kim, A., et al.\ 2019, A\&A, 632, A42
\bibitem[Bell et al.(2017)]{2017ASPC..509..303B} Bell, K.~J., Hermes, J.~J., Montgomery, M.~H., et al.\ 2017, 20th European White Dwarf Workshop, 303
\bibitem[Bell et al.(2016)]{2016ApJ...829...82B} Bell, K.~J., Hermes, J.~J., Montgomery, M.~H., et al.\ 2016, ApJ, 829, 82
\bibitem[Bell et al.(2015)]{2015ApJ...809...14B} Bell, K.~J., Hermes, J.~J., Bischoff-Kim, A., et al.\ 2015, ApJ, 809, 14
\bibitem[Bergeron et al.(2011)]{2011ApJ...737...28B} Bergeron, P., Wesemael, F., Dufour, P., et al.\ 2011, ApJ, 737, 28
\bibitem[Borucki et al.(2010)]{2010Sci...327..977B} Borucki, W.~J., Koch, D., Basri, G., et al.\ 2010, Science, 327, 977
\bibitem[Buchler et al.(1997)]{1997A&A...321..159B} Buchler, J.~R., Goupil, M.-J., \& Hansen, C.~J.\ 1997, A\&A, 321, 159
\bibitem[Camisassa et al. (2019)]{camisassa2019} Camisassa, M.~E., Althaus, L.~G., C{\'o}rsico, A.~H., et al.\ 2019,   A\&A, 625, A87
\bibitem[Chanmugam (1972)]{chanmugam1972} Chanmugam, G.\ 1972,   Nature Physical Science, 236, 83
\bibitem[Moya et al.(2018)]{2018A&A...620A.203M} Moya, A., Barcel{\'o} Forteza, S., Bonfanti, A., et al.\ 2018, A\&A, 620, A203
\bibitem[C{\'o}rsico et al.(2019b)]{2019arXiv191007385C} C{\'o}rsico, A.~H., De Ger{\'o}nimo, F.~C., Camisassa, M.~E., et al.\ 2019, arXiv e-prints, arXiv:1910.07385
\bibitem[C\'orsico et al. (2019a)]{corsico2019} C{\'o}rsico, A.~H., Althaus, L.~G., Miller Bertolami, M.~M., et al.\ 2019,   A\&ARv, 27, 7
\bibitem[C{\'o}rsico et al.(2012)]{corsico2012} C{\'o}rsico, A.~H., Althaus, L.~G., Miller Bertolami, M.~M., et al.\ 2012, A\&A, 541, A42
\bibitem[C{\'o}rsico et al.(2009)]{corsico2009} C{\'o}rsico, A.~H., Althaus, L.~G., Miller Bertolami, M.~M., et al.\ 2009, A\&A, 499, 257
\bibitem[C{\'o}rsico et al.(2008)]{corsico2008} C{\'o}rsico, A.~H., Althaus, L.~G., Kepler, S.~O., et al.\ 2008, A\&A, 478, 869
\bibitem[C{\'o}rsico et al.(2007a)]{corsico2007a} C{\'o}rsico, A.~H., Althaus, L.~G.,   Miller Bertolami, M.~M., et al.\ 2007, A\&A, 461, 1095
\bibitem[C{\'o}rsico et al.(2007b)]{corsico2007b} C{\'o}rsico, A.~H., Miller Bertolami, M.~M., Althaus, L.~G., et al.\ 2007, A\&A, 475, 619
\bibitem[Dziembowski(1982)]{1982AcA....32..147D} Dziembowski, W.\ 1982, Acta.   Astron., 32, 147
\bibitem[Faulkner \& Gribbin (1968)]{faulkner1968}   Faulkner, J., \& Gribbin, J.~R.\ 1968, Nature, 218, 734
\bibitem[Fontaine \& Brassard (2008)]{fon2008} Fontaine, G., Brassard, P.\ 2008 PASP, 120, 1043
\bibitem[Garc\'ia-Berro et al. (2010)]{garcia2010} Garc{\'\i}a-Berro, E.,   Torres, S., Althaus, L.~G., et al.\ 2010, Nature,   465, 194
\bibitem[Goupil et al.(1998)]{1998BaltA...7...21G} Goupil, M.~J., Dziembowski, W.~A., \& Fontaine, G.\ 1998, Baltic Astronomy, 7, 21
\bibitem[Harper \& Rose (1970)]{harper1970} Harper, R.~V.~R., \& Rose, W.~K.\ 1970, ApJ, 162, 963
\bibitem[Hermes et al.(2017)]{2017ApJS..232...23H} Hermes, J.~J., G{\"a}nsicke, B.~T., Kawaler, S.~D., et al.\ 2017, ApJS, 232, 23
\bibitem[Hermes et al.(2015)]{2015ApJ...810L...5H} Hermes, J.~J., Montgomery, M.~H., Bell, K.~J., et al.\ 2015, ApJL, 810, L5
\bibitem[Howell et al.(2014)]{2014PASP..126..398H} Howell, S.~B., Sobeck, C., Haas, M., et al.\ 2014, PASP, 126, 398
\bibitem[Kawaler, \& Bradley(1994)]{1994ApJ...427..415K} Kawaler, S.~D., \& Bradley, P.~A.\ 1994, ApJ, 427, 415
\bibitem[Koen, \& Laney(2000)]{2000MNRAS.311..636K} Koen, C., \& Laney, D.\ 2000, MNRAS, 311, 636  
\bibitem[Landolt (1968)]{landolt1968} Landolt, A.~U.\ 1968, ApJ, 153, 151  
\bibitem[Lasker \& Hesser (1969)]{lasker1969} Lasker, B.~M., \& Hesser, J.~E.\ 1969, ApJL, 158, L171
\bibitem[Lasker \& Hesser (1971)]{lasker1971} Lasker, B.~M., \& Hesser, J.~E.\ 1971, ApJL, 163, L89
\bibitem[Ledoux \& Sauvenier-Goffin (1950)]{ledoux1950} Ledoux, P.~J., \& Sauvenier-Goffin, E.\ 1950, ApJ, 111, 611
\bibitem[Luan, \& Goldreich(2018)]{2018ApJ...863...82L} Luan, J., \& Goldreich, P.\ 2018, ApJ, 863, 82
\bibitem[McGraw (1979)]{mcgraw1979} McGraw, J.~T.\ 1979, ApJ, 229, 203
\bibitem[Mestel (1952)]{mestel1952} Mestel, L.\ 1952, MNRAS, 112, 583
\bibitem[Miller Bertolami \& Althaus (2006)]{miller2006}   Miller Bertolami, M.~M., \& Althaus, L.~G.\ 2006, A\&A, 454, 845
\bibitem[Montgomery et al.(2019)]{2019arXiv190205615M} Montgomery, M.~H., Hermes, J.~J., \& Winget, D.~E.\ 2019, arXiv e-prints, arXiv:1902.05615
\bibitem[Ostriker \& Tassoul (1968)]{ostriker1968} Ostriker, J.~P., \& Tassoul, J.-L.\ 1968, Nature, 219, 577
\bibitem[Piotto(2018)]{2018EPSC...12..969P} Piotto, G.\ 2018, European Planetary Science Congress, EPSC2018-969
\bibitem[Ricker et al.(2015)]{2015JATIS...1a4003R} Ricker, G.~R., Winn, J.~N., Vanderspek, R., et al.\ 2015, Journal of Astronomical Telescopes, Instruments, and Systems, 1, 014003  
\bibitem[Robinson et al. (1976)]{robinson1976} Robinson, E.~L., Nather, R.~E., \& McGraw, J.~T.\ 1976, ApJ, 210, 211
\bibitem[Robinson et al. (1982)]{robinson1982} Robinson, E.~L., Kepler, S.~O., \& Nather, R.~E.\ 1982, ApJ, 259, 219
\bibitem[Rolland et al.(2018)]{2018ApJ...857...56R} Rolland, B., Bergeron, P., \& Fontaine, G.\ 2018, ApJ, 857, 56
\bibitem[Sauvenier-Goffin (1949)]{sauvenier1949}   Sauvenier-Goffin, E.\ 1949, Annales d'Astrophysique, 12, 39
\bibitem[Tassoul et al.(1990)]{1990ApJS...72..335T} Tassoul, M., Fontaine, G., \& Winget, D.~E.\ 1990, ApJS, 72, 335
\bibitem[Unno et al. (1989)]{unno1989} Unno, W., Osaki, Y., Ando, H., et al.\ 1989,   Nonradial oscillations of stars
\bibitem[Voss et al.(2007)]{2007A&A...470.1079V} Voss, B., Koester, D., Napiwotzki, R., et al.\ 2007, A\&A, 470, 1079
\bibitem[Walker et al.(2003)]{2003PASP..115.1023W} Walker, G., Matthews, J., Kuschnig, R., et al.\ 2003, PASP, 115, 1023
\bibitem[Warner \& Robinson (1972)]{warner1972} Warner, B., \& Robinson, E.~L.\ 1972, Nature Physical Science, 239, 2
\bibitem[Winget \& Kepler (2008)]{win2008} Winget, D. E., Kepler, S. O.\ 2008, ARAA 46,157
\bibitem[Winget et al.(1982)]{1982ApJ...252L..65W} Winget, D.~E., van Horn, H.~M., Tassoul, M., et al.\ 1982, ApJL, 252, L65
\bibitem[Winget(1982)]{1982PhDT........27W} Winget, D.~E.\ 1982, Ph.D. Thesis
\bibitem[Wu, \& Goldreich(2001)]{2001ApJ...546..469W} Wu, Y., \& Goldreich, P.\ 2001, ApJ, 546, 469  
\bibitem[York et al.(2000)]{2000AJ....120.1579Y} York, D.~G., Adelman, J., Anderson, J.~E., et al.\ 2000, AJ, 120, 1579
\bibitem[Zong et al.(2016a)]{2016A&A...585A..22Z} Zong, W., Charpinet, S., Vauclair, G., et al.\ 2016, A\&A, 585, A22
\bibitem[Zong et al.(2016b)]{2016A&A...594A..46Z} Zong, W., Charpinet, S., \& Vauclair, G.\ 2016, A\&A, 594, A46
\end{thebibliography}
\end{document}